\documentstyle[twocolumn,epsf,aps,tabularx]{revtex}
\draft
\begin{document}
\title{Spatial evolutionary prisoner's dilemma game with three strategies \\
and external constraints}

\author{Gy\"orgy Szab\'o,$^1$ Tibor Antal,$^2$  P\'eter Szab\'o,$^3$
and Michel Droz$^2$}

\address
{$^1$Research Institute for Technical Physics and Materials Science \\
P.O.Box 49, H-1525 Budapest, Hungary \\
$^2$Department of Theoretical Physics, University of Geneva,
1211 Geneva 4, Switzerland \\
$^3$Department of Ecology, J\'ozsef Attila University,
H-6721 Szeged, Egyetem u. 2, Hungary}

\address{\em \today}

\address {
\centering {
\medskip \em
\begin{minipage}{15.4cm}
{}\qquad  The
emergency of mutual cooperation is studied in a spatially extended evolutionary
prisoner's dilemma game in which the players are located on the sites of  cubic
lattices for dimensions $d=1$, 2 and 3. Each player can choose one of the
three following strategies: cooperation ($C$), defection ($D$) or  Tit for
Tat ($T$).  During the evolutionary process the randomly chosen
players adopt one of their neighboring strategies if the chosen
neighbor has higher payoff.  Moreover, an external constraint imposes that 
the players  always cooperate with probability $p$.
The stationary state phase diagram is computed by both using generalized 
mean-field approximations and Monte Carlo simulations. 
Nonequilibrium second order phase transitions associated with the
extinction of one of the possible strategies are found and the
corresponding critical exponents belong to the directed percolation
universality class. It is shown that forcing externally the collaboration 
does not always produce the desired result.
\end{minipage}
}}

\maketitle

\narrowtext

\section{INTRODUCTION}
\label{sec:intro}

Evolutionary game theory has attracted a lot attention during the past
years~\cite{smith,HS} in human sciences, political sciences, biology
and economics. In particular, the so-called evolutionary prisoner's
dilemma game (PDG), which is a metaphor for the evolution of
cooperation in populations of selfish individuals, has been minutely
investigated~\cite{smith,HS,sigmund,axelrod,weibull}. In the original
form of the PDG, only uniform populations with given strategies were
considered. However, it was realized~\cite{smith,NM} that new
interesting phenomena can occur when the PDG was expanded in such a way 
that local contests in a d-dimensional space could take place
(we shall use the abbreviation SPDG for such
systems). It turns out that these spatially extended models are
similar to the ones studied in nonequilibrium statistical
physics. They may exhibit cooperative behavior resulting in phase
transitions in the stationary state. Accordingly, it is very fruitful
to study SPDG like models using the tools developed in the framework
of nonequilibrium statistical physics.

In its simpler form, the PDG is a version of matrix games where the
symmetric incomes of the two players depend on their simultaneous
decisions, whether they wish to cooperate with the others or to defect.
Each player wants to maximize his individual income.  The highest
individual payoff (the temptation to defect) can be reached by the
defector against the cooperator receiving the lowest reward (sucker's
payoff). The mutual cooperation results in the highest total payoff
divided equally between the players. For mutual defections the players
get a lower payoff exceeding the value of sucker's payoff.  Two
rational players will both defect because this choice provides the
larger income independently of the partner's choice.

On the contrary, mutual cooperation dominates in economic and
biological systems where the contestants interact frequently.  In the
iterated round-robin PDG, the players, knowing the previous decisions,
have to choose between two options (defection and cooperation). For a
given round the contestants can be classified according to the total
individual payoffs they have obtained. Following the
Darwinian selection principle, at each round the worst player will
adopt the winner's strategy.

Extended numerical simulations have been performed to select the
``best strategy'' among many~\cite{axelrod,delahaye}. Best does not means that
this strategy will always win a fight against another strategy, but
means that it will obtain the highest payoff during a tournament
during which it will have to fight against many opponents having
different strategies. In~\cite{axelrod} the highest payoff was obtained by the
so-called ''Tit for Tat'' ($T$) strategy which cooperates in the first
round and then always repeats his co-player's previous decision. The
main characteristics of this strategy (never defect first, react to
the defection of the opponent and forgiveness) are crucial ingredients
to sustain the mutual cooperation against the defectors.
In particular, extensive simulations (see~\cite{sigmund} for a summary)
for the case in which the players adopt one of the two following strategies: $C$
(cooperate unconditionally) or $D$ (defect unconditionally) have shown that
the cooperators were disappearing in the stationary state. The introduction of
some $T$'s strategies has an important effect. For short times, the D's population increases while the C's one decreases, leading to the decrease of 
D's payoffs. As a consequence, the T's can invade the D's population.

In order to study the spatial effects Nowak and May~\cite{NM} have
introduced an SPDG consisting of a two-state cellular automaton. The
players are located on a regular lattice in a d-dimensional
space and can adopt the C's or D's strategy. Each
player is fighting with the individuals belonging to a given
neighborhood. The player's strategies are upgraded simultaneously in
discrete time steps according to the following rule: each player
adopts the best strategy found in his neighborhood. This model
exhibits a rich variety of spatial and temporal patterns as a
function of the payoff $b$ characterizing the temptation to defect.
Other SPDG models have also been investigated \cite{kiluno,kildue,kiltre}.
In particular Killingback and Doebeli \cite{kilquatro} have shown that
``Pavlov'' like strategies can be even more efficient than
``Tit for Tat'' in some circumstances.

Nowak {\it et al.} \cite{NBM} have also extended the above analysis by
allowing stochasticity (irrational choices) during the evolutionary
process. The degree of stochasticity is governed by a parameter called 
$m$, and in the limit $m \to \infty$, one recovers the deterministic case.
According to the value of $b$, several stationary states are
possible.

In some related models~\cite{epdg2s}, Szab\'o and  T\H oke have shown that
the different stationary state phases were separated by second order 
nonequilibrium phase transitions lines.  The associated critical 
exponents belongs to the directed percolation (DP) universality 
class\cite{epdg2s,CO}.

Both the importance of the presence of the T's in the PDG and the richness 
associated with the spatially extended aspect motivated us to study a new class of SPDG. In the present work, we study a novel aspect of the SPDG, namely what is
happening when the cooperators are enforced by some external
constraints.  More explicitly, we consider a SPDG with the three
strategies $D$, $C$, and $T$ and investigate the effects of random
adoption (or forcing) of $C$ strategies. This new effect can be 
interpreted as an
attempt by a government or by any other organization to enforce the
cooperation among individuals by forcing some of them, chosen randomly
with a given probability $p$, to cooperate. A different interpretation 
can also be given. Among the
player's community, some old players  resign and are replaced by younger 
ones having a different educational background making them more open to 
collaborate. As we shall see the
effects of this external constraint are rather surprising. Indeed, in
dimensions 1,2 or 3, the presence of the constraint reduces the
cooperation if $p$ is less than a given threshold value $p_c$
depending upon the dimensionality of the system.  The cooperation is
enforced only if the constraint is strong enough, i.e.  if $p>p_c$.

The nonequilibrium second order phase
transitions describing the extinction of one of the possible 
strategies are found to belong to the directed percolation universality class.

The paper is organized as follows.  The model is defined in section
II. Its properties are analyzed in  mean-field
approximation in section III. The properties of  the model without 
constraint are discussed in section IV.
The model with constraint is studied in section V, both in mean-field 
approximation and by Monte Carlo (MC) simulations, for respectively
1, 2 and 3 dimensions. Finally, conclusions are drawn in section VI.

\section{THE MODEL}
\label{sec:model}

We consider a SPDG model in which the players are located on the sites
of a $d$-dimensional cubic lattice of linear size $L$ where periodic
boundary conditions are assumed.
Each player can adopt one of the
three following strategies $D$ (defects), $C$ (cooperates) or $T$ (Tit
for Tat) and interacts with its $2d$ nearest neighbors. The total
payoff of a given player is the sum of the payoffs coming from the
interaction with all its nearest neighbors. We use an extension 
(including the Tit for Tat strategy) of the payoff matrix used by 
Nowak and May
\cite{NM}. The individual payoffs for the players $P_1$
and $P_2$ as a function of their strategies are given in Table 1. The
only free parameter $b$ ($1<b<2$) measures the temptation to defect.
Note that the above payoff matrix does not take into account that the $T$
players always try to cooperate with D's during the first round. Thus these values are  considered as the averaged (or in this present case 
stabilized) payoffs. 
\begin{center}
\begin{tabular}{|c|c|c|c|} \hline
$P_1$ $\backslash$ $P_2$ & $D$ & $C$ & $T$ \\ \hline $D$ & 0
$\backslash$ 0 & $b$ $\backslash$ 0 & 0 $\backslash$ 0 \\ \hline $C$ &
0 $\backslash$ $b$ & 1 $\backslash$ 1 & 1 $\backslash$ 1 \\ \hline $T$
& 0 $\backslash$ 0 & 1 $\backslash$ 1 & 1 $\backslash$ 1 \\ \hline
\end{tabular}
\end{center}
\vspace*{1mm}
Table 1: Payoff matrix of the model
\vskip 0.3truecm

It is legitimate to use this payoff matrix providing that the strategy adoption
is rare comparing to the frequency of the game. It makes the simulation simpler and more efficient.
Notice that similar payoff matrices can be obtained when substituting some
other ``nice'' strategy  (as Pavlov which cooperates in the first round for example, a strategy which is nevertheless less efficient than T in this case)~\cite{sigmund,axelrod} for $T$.

For the non constrained case ($p=0$), the system evolves in discrete
time according to the following MC process. Starting from a
random initial state, a site is chosen randomly. This site updates its
strategy by first selecting randomly one of its nearest neighbor 
and second, by adopting the strategy of this nearest neighbor 
only if  it is having a higher payoff.  
A MC steps consists in updating each lattice site once, on the average.

In the constrained case, at each time step, each player is forced to
adopt the cooperative strategy $C$ with a probability $p$.

Thus the dynamics of the constrained model is the following.
\begin{itemize} 

\item One chooses randomly one player.

\item With probability $p$ this player adopts the $C$ strategy.

\item With probability $(1-p)$ the player searches for a better 
strategy according to the procedure described above.

\item The players update their payoffs.

\end{itemize}

The  model is characterized by three free parameters: $b,p,d$. 

It is hopeless to find exact analytical solutions for such models.
Accordingly, we shall first study them in the framework
of  mean-field like approximations and then investigate them by numerical
simulations.

\section{MEAN-FIELD LIKE APPROXIMATIONS}
\label{sec:MFA}

The simplest mean-field approximation consists in neglecting all the
spatial correlations in the systems. This amount to consider a model
in which for each player, the partners are chosen randomly in the
system instead of being restricted to a particular neighborhood. Each
player interacts with the same number of counterparts. The
dimensionality of the system plays no role. The simplest mean-field like
approximations have been successfully used previously for similar
problems and more details concerning this approximative scheme can be
found in textbooks~\cite{smith,HS,weibull}.

Within this approximation, the dynamics of the system is completely
described by the time dependent concentrations:
\begin{equation}
c_{\alpha}(t) = \langle N_{\alpha}(t) \rangle / L^d, \quad
(\alpha=~C,D,T)
\end{equation}
where $N_{\alpha}(t)$ is the number of players with strategy $\alpha$
at time $t$. These concentrations satisfy the normalization condition
$c_D+c_C+c_T=1$.

According to Table 1 the average payoffs for each strategy are:
\begin{equation}
m_D=b c_C \ ,\ \ m_C=c_C+c_T \ ,\ \ m_T=c_C+c_T \ .
\end{equation}
Notice that the $C$ and $T$ strategies have the same payoffs,
therefore no strategy exchange will occur among them.

Following the evolutionary rules given in Sec.~\ref{sec:model}, the
concentrations $c_{\alpha}(t)$ obey the following equations of motion:
\begin{eqnarray}
\dot{c}_D&=&-p c_D \mp (1-p) c_D(c_C+c_T) \ , \nonumber \\
\dot{c}_C&=&+p (c_D+c_T) \pm (1-p) c_D c_C \ ,
\label{eq:mfa} \\
\dot{c}_T&=&-p c_T \pm (1-p) c_D c_T \ , \nonumber
\end{eqnarray}
where the upper and lower signs refer respectively to the cases when
strategy $D$ is dominated by $C$ and $T$ ( $m_D < m_C=m_T$), and when
$D$ dominates $C$ and $T$ ($m_D > m_C=m_T$).

\begin{figure}
\centerline{\epsfxsize=6.5cm
            \epsfbox{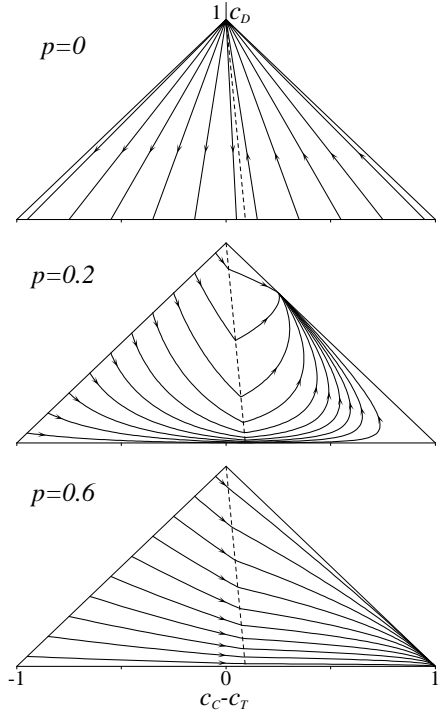}
            \vspace*{1mm} }
\caption{Trajectories in the two-dimensional phase space for three
different values of $p$ as indicated in the figures. The dashed lines
divide the phase spaces into two regions: on their left (right) hand
side the payoff of the $D$ strategy is lower (higher) than those of
the $C$ and $T$ ones.}
\label{fig:MFAARR}
\end{figure}

The numerical integration of the above equations of motions leads,
for several values of $p$, to
the flow diagrams shown in Fig.~\ref{fig:MFAARR}.

The quantities represented on the  vertical and horizontal axes are 
respectively 
$c_D$ and $c_C-c_T$. The upper corner of the triangle corresponds to the state
of $c_D=1$, while the lower left and right corners describe
respectively homogeneous states with $c_T=1$ and $c_C=1$.

For $0<p<1/2$ the stationary state solution of the above equations  is:

\begin{equation}
c_D={1-2p \over 1-p};\ \ \ c_C={p \over 1-p};\ \ \ c_T=0
\label{eq:MFsol}
\end{equation}
while for $p>1/2$ the system goes to the adsorbing state ($c_C=1$ and
$c_D=c_T=0$).  Surprising the $T$ strategy extincts if $p > 0$.

Without constraints, i.e. for $p=0$, the system tends  either to a
homogeneous $D$ adsorbing state ($c_D=1$) or to a mixed state of $C$
and $T$ strategies ($c_D=0$) depending on the initial
conditions.
Note that the above stationary states are independent of the value of
$b$. Strikingly different behaviors will be observed beyond this simplest
mean-field approximation when the local fluctuations and  short-range correlations are taken into account as we shall see in the
following sections.

More elaborated mean-field like approximations can be devised.
The basic idea is to take explicitly into account some of the spatial 
correlations by computing the probability of appearance of all the 
possible configurations of a small cluster containing $n$ sites. In one dimension, 
one considers  a cluster of $n$ consecutive lattice sites.
The equations of motion for these  probabilities  follows from the evolution rules
of the system. Details concerning the one-dimensional case are given 
in~\cite{GMF1D}. Note that already the one site mean-field approximation 
($n=1$) differs from the simple mean-field solution given above.
In higher dimensions pair or square mean-field like approximations have 
been used, and a detailed description of these methods
can be found in  \cite{epdg2s,ST}. Note that now, the predictions of these 
approximations depends upon the dimension of the system. Accordingly, the cases $d=1,2$ and $3$ will be discussed in different subsections.

\section{MODEL WITHOUT EXTERNAL CONSTRAINT}
\label{sec:without}

Let us start by considering the case with no external constraint, i.e. $p=0$.
In this case, the dynamic is simple: a given player
adopts the strategy of one of his randomly chosen neighbor
providing that  this neighbor has a higher payoff. As we shall see,
the only stationary states are either always trivial cooperator like 
($C$ or $C$-$T$) or a pure defector one ($D$).

When  the random initial state is
made of only cooperators and defectors, one founds that, both in  
$d=1, 2$ and $3$ dimensions, the stationary state of the system is 
a pure defector one.

The reason in $d=1$ is simply that a $D$ player has always a 
higher payoff than a neighboring $C$ player.

In $d=2$ several configurations should be analyzed. 
The most favorable situation for $C$ to win is when it is adjacent
to a flat $C$-$D$ interface. However, if this interface
has an irregularity then  the $D$'s can invade the sea of $C$ players.
Indeed, a new born $C$ in the $D$ half-plane is always weaker than the $D$'s
at the interface and thus, such $C$'s disappears sooner or later.
However, at a given time and with some finite probability, 
two next nearest neighbor $C$ players could be present in the $D$ half-plane. 
It results a payoff $3b$ for the $D$ player squeezed between three $C$ and this $D$ can now invade  the 
$C$'s and sweep off one layer of them independently of the value of $b$.

The $d=3$ case is very similar, but the process is slower. Indeed,
four nearest neighbors of a $D$ player (called $D_1$) are necessary to be 
invaded by the $C$'s if the value of  $b$ is  close to 1. Once it has 
occured, $D_1$ is strong enough to cross the interface and then 
destroys all the $C$'s.

For a large system with initially a finite proportion of $T$ players, 
the stationary state is always a cooperator like state ($T$-$C$).
This asymptotic behavior can be easily understood in the one dimensional case.
Let us suppose that there are four neighboring $T$ ($TTTT$) in 
the initial state of a 1d system, 
(this is practically always the case in a sufficiently large system 
with a finite probability to have some $T$'s in the initial state),
then they kill all of the $D$'s.
Even in the worst case two $D$'s could invade into to the $T$ population, 
($CDDTTDDC$), but then the central $T$'s become stronger (payoff=1) 
then the $D$'s (payoff=0). The $D$'s kill all of the neighboring $C$'s,
and even these two $T$'s can invade the whole $D$ area.

This argument can be extended to $d$ dimensions. A cube of linear size 4, made 
of $T$ players is enough to guarantee the cooperation in the stationary 
state. The reason is that the $D$'s surrounding this cube cannot destroy 
the central $T$ players inside a cube of linear size 2. Indeed, the 
central $T$'s have always $d$ $T$'s neighbors to cooperate with (payoff=$d$), 
but the neighboring $D$'s can  only have one $C$ neighbor 
(payoff$\le b<d$ if $d\ge 2$).

\section{MODEL WITH EXTERNAL CONSTRAINT}
\label{sec:noc}

We now consider the case in which the players are located on the sites of
a d-dimensional cubic lattice in the presence of an external constraint
imposing to each player to choose the $C$ strategy with probability $p$.
The cases $d=1,2$ and $3$ will be considered.

\subsection{One-dimensional system}
\label{subsec:1D}

In the one-dimensional model the players located on the sites of a chain
interact with their two nearest neighbors. It is easy to see that the dynamics 
is independent of the value of the parameter $b$ in its domain of definition.

\begin{figure}
\centerline{\epsfxsize=7cm
            \epsfbox{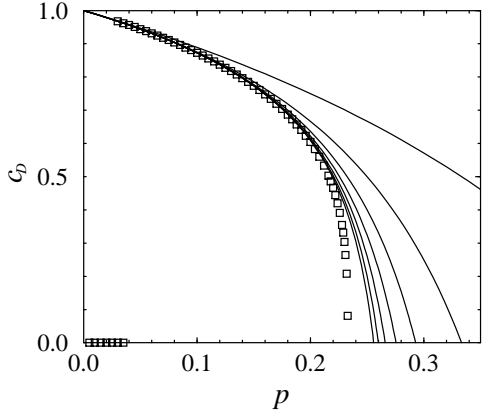}
            \vspace*{1mm}       }
\caption{The concentration of strategy $D$ as a function of $p$
in the one-dimensional system. The open squares are the MC data
obtained for $L=16384$. The solid lines represent the results of
$n$-point approximations ($n=1,...,7$ from right to left).}
\label{fig:CP1D}
\end{figure}

A systematic MC analysis of the stationary states was performed by
varying the value of $p$ for different system sizes between
$L=32$ to $L=16384$. Our simulations show that the T's strategy extincts for all values of $p$. However, as a 
function of $p$, the stationary state 
can be either a symbiosis of $C$ and $D$ strategies or a pure $C$ 
state as shown in Fig.~\ref{fig:CP1D}.
For very small values of $p$, one has a pure $C$ stationary state 
($c_C=1$) but, when $p$ reaches the value $p_1$, a first transition 
occurs to a stationary state in which the two strategies $D$ and $C$ coexist.
At  $p=p_{c2} >p_{1}$ the system undergoes a second continuous transition  
from the $C$-$D$ stationary state to the pure absorbing $C$ state.

The transition at $p_{c2}$ is easy to understand. 
In one dimension and when only $C$ and $D$ strategies are present
our model is equivalent to the contact process (CP)
which was originally introduced as 
a simple model for epidemics~\cite{Har74}.
In the CP a particle (sick person) can disappear at rate 1,
and an empty site (healthy person) can become occupied 
with rate $\lambda z/2$,
where $\lambda$ is the control parameter and $z$ is the
number of particles in the neighborhood of the empty site.
In our model the $D$ strategy, which in $d=1$ is always better
than the $C$ one, plays the role of the sick persons
and the $C$ strategy, which can only be created in the system 
through the constraint, corresponds to the healthy individuals.

Starting from a $C$-$D$-$T$ initial state and after the extinction of 
$T$'s the possible stationary states are the pure $C$ or a $C$-$D$ states.
The $C$-$D$ state  corresponds to the steady state of the CP as well.
For $p$ around $p_{c2}$, the $T$'s disappear rapidly and thus do not affect
the extinction of the $D$'s. Hence the second transition to the absorbing 
state corresponds to the CP's one. This transition occurs at 
$p_{c2}=1/(1+\lambda_c)=0.23267$ \cite{Jen93} and is believed to 
belong to the DP universality class~\cite{DPconj,dp}. 

\begin{figure}
\centerline{\epsfxsize=7cm
            \epsfbox{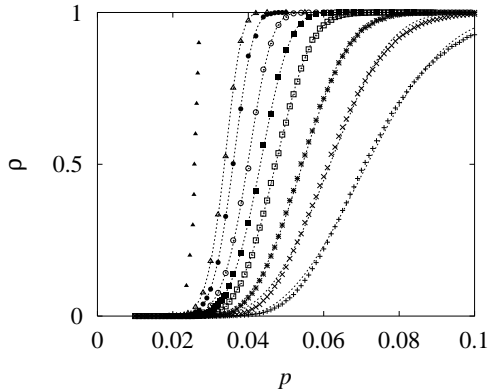}
            \vspace*{1mm}       }
\caption{The probability, $\rho_L(p)$, of reaching the $C$-$D$ state for
different system sizes (from right to left $L=50$, 100, 200, 500, 1000,
2000, 5000, 10000 and the interpolation for $L=\infty$). The dotted
lines represent a fit described in the text.}
\label{fig:rat}
\end{figure}

The behavior of the first transition at small $p$'s is much less clear.
It turns out that for our model two characteristic parameters 
$p_{\alpha}(L)$ and $p_{\beta}(L)$, depending on the system size $L$, 
can be introduced.
For $p<p_{\alpha}(L)$ the stationary state is 
always the pure $C$ state while for $p>p_{\beta}(L)$ the system evolves
to a $C$-$D$ coexistence phase, which is a steady state of the CP.
For $p_1(L)<p<p_2(L)$,  the system can evolves towards one of the two
possible stationary states, depending on the particular realization 
of the random numbers and on the initial state.

\begin{figure}
\centerline{\epsfxsize=7cm
            \epsfbox{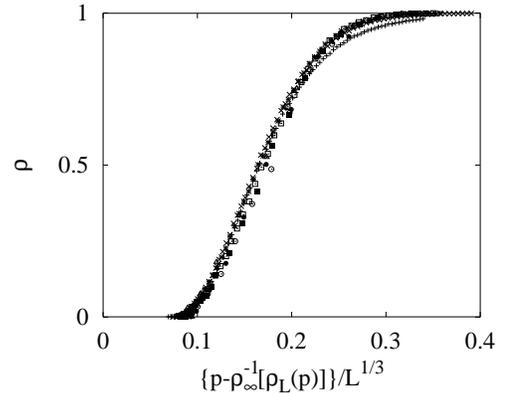}
            \vspace*{2mm}       }
\caption{The probability $\rho$ as a function of the scaled parameter.
The symbols represents the different system sizes as in
Fig.~\ref{fig:rat}.} 
\label{fig:ratscal}
\end{figure}

The probability, $\rho_L(p)$, of reaching the $C$-$D$ state
has been investigated numerically. The data are plotted in Fig.~\ref{fig:rat}
and can be well fitted by a function of the type 
$1-\exp(-c_1(p-p_0)^{c_2})$, where $c_1$, $c_2$ and $p_0$ are fitting
parameters.
The  limiting function  $\rho_{\infty}(p)$ was obtained by using usual 
finite size scaling methods. As shown in Fig.~\ref{fig:ratscal}, 
the functions $\rho_L(p)$ collapse on a single curve if one plots 
$\rho$ as a function of $\{p-\rho^{-1}_{\infty}[\rho_L(p)]\}/L^{1/3}$.
This shows that $\rho_{\infty}(p)$ do not collapse to zero, hence the 
phase transition takes place at a finite value of $p \simeq 0.025$.

\begin{figure}
\centerline{\epsfxsize=7cm
            \epsfbox{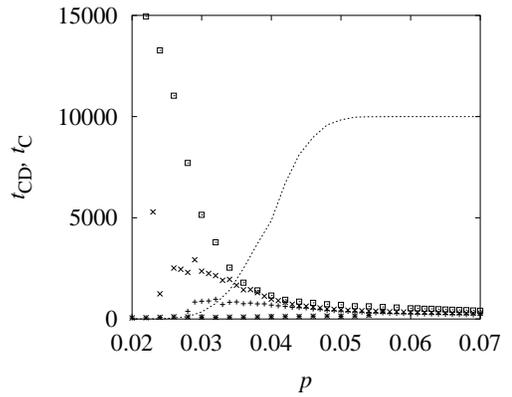}
            \vspace*{2mm}       }
\caption{The respective times $t_C$ ($L=2000 (*)$) 
and $t_{CD}$ ($L=200$ (+), 500 ($\times$), 2000 ($\Box$)) needed
to reach the $C$ and the $C$-$D$ stationary states. For comparison,
$10000 \times \rho_{L=2000}(p)$ is also presented with dotted line.}
\label{fig:times}
\end{figure}

The respective times $t_C$ and $t_{CD}$ needed to reach the pure $C$ and
the $C$-$D$ stationary states have also been investigated.
Both times show a singular behavior (see Fig.~\ref{fig:times}).
Unfortunately, we were not able to find a reasonable scaling fit for those
data. However, it is obvious from the simulations,
that $t_{CD}$ exhibits a much stronger singularity than $t_{C}$.
Hence, for a large system, the time needed to reach the $C$-$D$ state is much 
larger than the one needed to reach the $C$ state, even if the system 
evolves to the $C$-$D$ state more frequently. 

Beside the MC simulations, the properties of the system were also
investigated using  generalized mean-field approximations. Assuming the
coexistence of all the three strategies these calculations can be performed
numerically on clusters of sizes as large as $n=4$. Contrary to the MC
results (see Fig.~\ref{fig:CP1D}) these approximations predicted the existence
of a $C$-$D$-$T$ state at small $p$ values. As far as the T's are concerned,
we observed that the maximum value of $c_T$ is decreasing when the cluster sizes
$n$ were increasing, providing us a trend for the extinction of the T's.

The contradiction between the present mean-field and MC
results refers to the importance of (interfacial) invasion phenomena
detailed later on.

Allowing only two strategies ($C$ and $D$) the above mean-field analysis
can be performed for larger cluster sizes up to $n=7$. The results 
are compared with the MC ones in Fig.~\ref{fig:CP1D}. As expected,
the accuracy of the generalized mean-field method increases with the
cluster size $n$. The extrapolation of the results obtained for finite
values of $n$ ($n=1,...,7$) leads to a critical value
$p_{c2}^{(MF)} \simeq 0.235$. The quality of this approximative scheme
can be estimated by comparing the value of $p_{c2}^{(MF)}$ with the best
known numerical value $p_{c2}=0.23267$ ~\cite{Jen93}.

\begin{figure}
\centerline{\epsfxsize=8cm
            \epsfbox{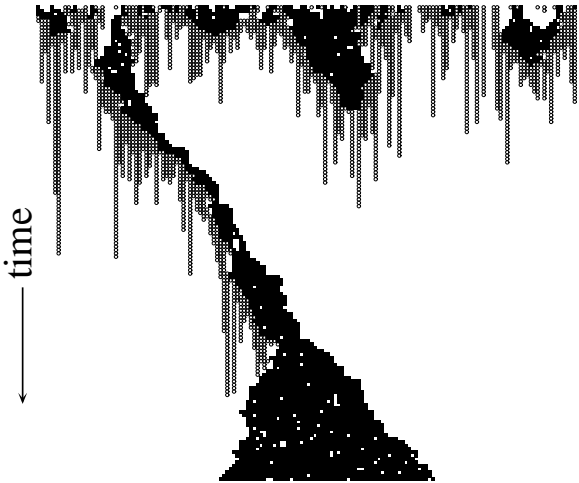}
            \vspace*{1mm}       }
\caption{Evolution of strategy distribution for $p=0.04$
starting from a random initial state. The $k$-th row
shows the positions of the $D$ (closed squares), $C$ (empty), and $T$ 
(open circle) strategies at the time $2k$ measured in MC
steps.}
\label{fig:EVOL1D}
\end{figure}

In order to understand the behavior of the model around the first
transition it is interesting to examine the time evolution of the
system in its transient regime. As illustrated in Fig. \ref{fig:EVOL1D},
one can observe a domain growth process controlled by cyclic invasions 
\cite{FKB}. This picture suggests that the most relevant aspect of the
dynamics is the collision between the  fronts separating different strategies.

Let us first investigate the motion of a front separating a cluster of 
$C$ from a $D$ one.
The $D$ domains contain some $C$ sites coming from the external constraint. 
However, the life times of the $C$'s are very short [$\sim (1-p)$]. 
Within a $D$ domain the average $C$ concentration is equal to that of 
the CP and can be well approximated  by the simple mean-field approach 
as we have seen above. The $D$ strategy invades the territory of the 
$C$ one (absorbing state). Due to the reflection symmetry the
invasion front can move to the left or to the right with the same absolute 
value of the velocity.
To first order in $p$ in the limit $p \to 0$, the absolute value of the 
average invasion velocity can be estimated by using the configuration 
probabilities given for $n=1$. This calculation yields 
\begin{equation}
v = {1-3 p \over 2} \ .
\label{eq:vdc}
\end{equation}

\begin{figure}
\centerline{\epsfxsize=7cm
            \epsfbox{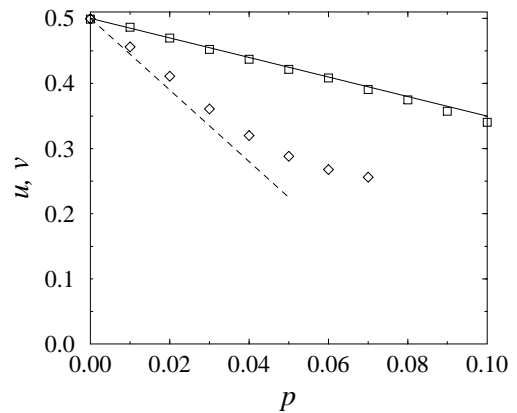}
            \vspace*{1mm}       }
\caption{Velocities of $D \to C$ (open squares) and $T \to D$ 
(diamonds) invasions as a function of $p$. The solid and dashed lines
indicate the analytical predictions in linear approximation.}
\label{fig:vu}
\end{figure}

Let us now consider the invasion of the $T$ strategy into the $D$ one.
Inside a territory which has been invaded by the $T$ strategy, the $C$
strategy is setting up with a probability $p$. As both $C$'s and $T$'s
have the same payoff, no adaptation of strategies occur between them.  
Assuming that this invading front travels with a constant velocity $u$,
the probability of having a $C$ player at a site $k$ steps behind the front
is $1-e^{-p \tau (k)}$. $\tau (k) = (k+1/2)/u$ is the averaged time 
it takes to the front to move over a distance of $k$. Thus, it follows that,
to leading order in $p$,
\begin{equation}
u = {1-11 p \over 2} \ .
\label{eq:vtd}
\end{equation}
We note that above approximations lead to a velocity for the $D \to C$ front 
which is larger than the one for the $T \to D$. This prediction can be
compared with the results of the MC simulations (see Fig.~\ref{fig:vu}). 
For the $D \to C$ front, the agreement is very good if $p < 0.1$, while  
for the $T \to D$ case, the approximation reproduces well only the linear 
part near the origin.

Another consequence of the external constraint concerns the
life-time of the $T$ clusters. According to the above picture, the probability
that a $T$ cluster dyes out during the time evolution can be approximated by:
\begin{equation}
P_T=\prod_{k=0}^{\infty} \left[ 1- \exp(-p (k+1/2)/u) \right] \ .
\label{eq:prod}
\end{equation}
The inverse of $P_T$ can be interpreted as the average life time $\tau_T$ 
of an invading $T$ cluster.
Substituting an integral for the infinite sum appearing in the logarithm 
of the above expression leads to:
\begin{equation}
\tau_T \simeq \exp({\pi^2 u \over 6 p}) = \exp(\pi^2 { 1-11p \over 12 p}) \ .
\label{eq:Q}
\end{equation}
This expression shows a fast increase of the life-time when $p$ is
decreasing. For example, a $T$ cluster dyes out in about $10^5$ MC
steps (MCS) if $p=0.04$. This estimated life-time is significantly
larger than those found by MC simulations [$\tau_T^{(MC)}(p=0.04)
\approx 800$ MCS] as indicated in Fig.~\ref{fig:EVOL1D}. The $p$-dependence
of the MC data can be well approximated by the function
$\tau_T = 3.26 \exp (0.226/p)$ within the region $0.025 < p < 0.08$ we could
study. The large discrepancy between the mean-field and MC
results refers to the enhanced role of the velocity fluctuations. 

Starting from a random initial state spatially separated, domains are 
rapidly formed. Then two different situations can occur.
First, $T$ clusters are present one both ends of each 
$D$ clusters leading to a fast extinction of $D$'s.
Accordingly, the system  evolves to a pure $C$ state. 
Second, after some time the system reaches a state characterized by the 
presence of only one $D$ cluster, having a $T$ island at only one of its 
ends, in a sea of $C$. As the $D \to C$ invasion is faster than the $T \to D$,
the $T$'s can never destroy the $D$'s and due to the finite life-time of the 
$T$ cluster the stationary state is a $C$-$D$ coexisting one.
The exponential increase of the life-time of this $T$ cluster as $p \to 0$
explains the singular behavior observed in $t_{CD}$.

For $p<p_1(L)$ the life-time of the $T$ players are so long
that all the $D$ clusters are surrounded by $T$'s. Accordingly,
the first scenario described above is always present, leading to a pure 
$C$ state.
In the contrary, for $p>p_2(L)$ the short lifetime of $T$'s
insures that the $T$'s disappear  rapidly, allowing for the growth of the 
$D$ clusters. As a consequence the stationary state is a $C$-$D$ one.

\subsection{Two-dimensional system}
\label{subsec:2D}

The players are located on the sites of a two-dimensional square lattice.
According to the payoff matrix (see Table 1), two ranges of values of $b$
have now to be distinguished, namely, $1<b<3/2$ and $3/2<b<2$. For any
value of $b$ in one of those ranges the dynamics is the same.

\begin{figure}
\centerline{\epsfxsize=7cm
            \epsfbox{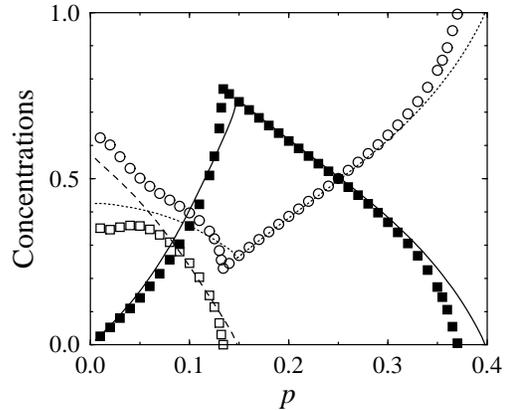}
            \vspace*{1mm} }
\caption{Stationary state densities of $D$ (closed squares), $C$ (open
circles) and $T$ (open squares) strategies as a function of $p$ as
obtained by MC simulation of a system of size $512 \times 512$ for
$3/2<b<2$. The solid ($D$), dotted ($C$) and long-dashed ($T$) lines
indicate the predictions of the square mean-field like approximation.}
\label{fig:CP2D}
\end{figure}

Let us first consider the simulations performed for the value $3/2<b<2$.
The situation is summarized in Fig.~\ref{fig:CP2D}, in which the stationary
values of the strategy concentrations are plotted as a function of $p$.
For $p < p_{c1}^{(MC)} = 0.1329(1)$,
the system reaches a stationary state in which the three strategies
$C,D$ and $T$ coexist. Here it is worth mentioning, that for small $p$
values ($p \alt 0.03$) the finite system can reach the absorbing state if
the initial state has been chosen randomly. Below this size-dependent
threshold value the three-strategy state can be formed and sustained 
by slow decreasing of $p$ during the simulation. In this case the extinction
of $T$'s and $D$'s is a consequence of fluctuations (detailed later on)
and the  coexistence of the three strategies is considered as the real
stationary state in the limit $L \to \infty$.
The simulations performed for $p \ge 0.005$ show that $c_D$ decreases
linearly with $p$ when $p \to 0$. 

When $p_{c1}^{(MC)}< p <p_{c2}^{(MC)}=0.3671(1)$ only the strategies
$C$ and $D$ survive. Finally, when $p>p_{c2}^{MC}$ the system reaches a pure
$C$ absorbing state. 

The phase diagram obtained by numerical simulations can be compared
with the ones obtained using the extended mean-field approximation
described in Sec.~\ref{sec:MFA}. At the level of pair approximation,
one finds $p_{c1}^{(pair)}=0.1704$ and $p_{c2}^{(pair)}=0.4236$,
while for the square mean-field approximation one finds
$p_{c1}^{(sq)}=0.1482$ and $p_{c2}^{(sq)}=0.3980$. These latest
results are plotted in Fig.~\ref{fig:CP2D}. They compare well with the
MC values given above.

\begin{figure}
\centerline{\epsfxsize=7cm
            \epsfbox{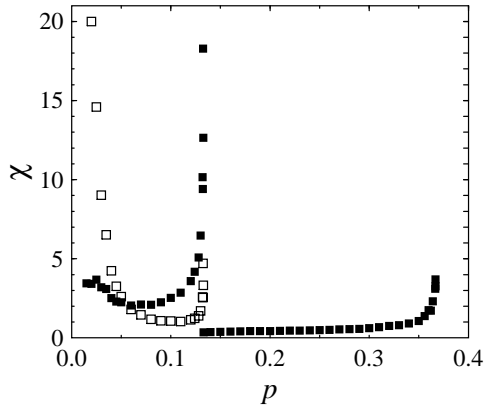}
            \vspace*{1mm} }
\caption{Concentration fluctuations for the $D$ (closed squares) and T
(open squares) strategies for a $512 \times 512$ system.}
\label{fig:FLUC2D}
\end{figure}

Some complementary information can be obtained by studying the concentration 
fluctuations defined as:
\begin{equation}
\chi_{\alpha}=L^d \langle(N_{\alpha} / L^d -c_{\alpha})^2\rangle,  \quad
(\alpha=~C,D,T) \ .
\label{eq:chi}
\end{equation}
When $p \to 0$, the concentration fluctuations $\chi_C$ and $\chi_T$ are
diverging while $\chi_D$ remains regular as shown in Fig.~\ref{fig:FLUC2D}.
However, the sizes of the systems investigated were
not large enough to conclude that $\chi_C$ and $\chi_T$ are diverging as
power laws of $p$ in the limit $p \to 0$.

Moreover, the simulations
suggest that for the stationary state in which the three strategies
coexist, the typical size of $T$ domains (as well as their persistence
time) is proportional to $1/p$.  Similar behavior was found in the
forest fire models \cite{FF} introduced by Bak {\it et al.} \cite{BCT}
to study the self-organized criticality.

The points $p_{c1}$ and $p_{c2}$ are critical points where a second
order nonequilibrium phase transition takes place. Indeed, the $T$
concentration vanishes at $p_{c1}$ as $c_T \sim (p_{c1}-p)^{\beta_1}$,
while the $D$'s concentration vanishes at $p_{c2}$ as $c_D \sim
(p_{c2}-p)^{\beta_2}$. In order to justify this behavior a very careful
numerical analysis was performed, using longer sampling times in
the vicinity of critical points. Fitting the numerical data leads to
the above mentioned values of the critical points and
$\beta_1=\beta_2=0.57(3)$ which is compatible with the 
directed percolation exponent  as expected on general ground\cite{dp}. This fact
is confirmed by the study of concentration fluctuations defined by
Eq.~(\ref{eq:chi}). Sharp increases of the concentration fluctuations are
expected at second order phase transitions. Fig.~\ref{fig:FLUC2D}
illustrates this point. The concentration fluctuations of $C$, $D$ and
$T$ strategies behave, near the transition
point $p_{c1}$, as $\chi \sim (p_{c1}-p)^{-\gamma_1}$ with
$\gamma_1=0.37(6)$. A similar behavior was found at the
transition point $p_{c2}$ with an exponent $\gamma_2=0.37(9)$. These
values are very close to the one of directed percolation: $\gamma_{DP}
\approx 0.35$ in two dimensions\cite{dp}.

The above data suggest that $p=0$ is another critical point. However, for 
the reasons explained previously, it was not possible to extract reliable 
exponents.

\begin{figure}
\centerline{\epsfxsize=8cm
            \epsfbox{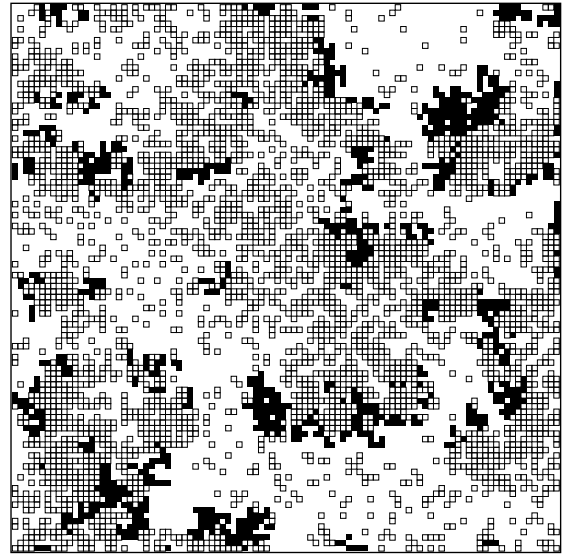}
            \vspace*{1mm} }
\caption{Distribution of defectors (black boxes), cooperators (white
area) and Tit for Tat strategies (open squares) in a system of size
$100 \times 100$, subset of a larger system of size $256 \times 256$ for  
$p=0.04$.}
\label{fig:GANGS}
\end{figure}

It is interesting to analyze how the three strategies coexist for
small values of $p$. As an example, let us consider the snapshot of
the stationary state of a system with $p=0.04$ (see Fig.~\ref{fig:GANGS}). 

One can observe that ''dark areas'' (made of defectors) are invading
the territory (''white areas'') of cooperators, simultaneously,
the ''dark areas'' are invaded by the $T$ actors (open squares).  
However, the domination of the $T$'s is prevented by the external 
constraint which
is leading to the growing of $C$ areas within the $T$
territory.  When $p$ decreases, the $T$ territory expands while the
growth of white areas slows down. The dark islands become sparse.
On the contrary, an increase of the value of $p$ accelerates the
spreading of the $C$ areas as well as their occupation by the
defectors.  Consequently, the $D$ population increases with $p$
and the number of $T$ competitors decreases and vanishes at $p=p_{c1}$.

As a result of these cyclic dominant processes a self-organized domain
structure is maintained in the system. Analogous
spatio-temporal structures have already been observed by Satulowsky
and Tom\'e in a two-dimensional predator-prey system \cite{ST} and by
Tainaka and Itoh when studying competing species \cite{TI}. Both 
models belong to the family of spatially-extended Lotka-Volterra models
predicting oscillatory behavior of concentrations in the simple
mean-field approximation\cite{LV}.

One can also analyze the evolution of the typical strategy configurations
in the vicinity of the phase transition taking place at ($p=p_{c1}$). One
recognizes isolated colonies of the $T$ strategies whose motions remind us 
of the branching annihilating random walk models. It is known that this
model belongs to the DP universality class too \cite{BARW}.

The rate of mutual cooperation is related to the average payoff per site.
The maximum average payoff (which value is 4) is reached when all the players
cooperate with their neighbors. On the other hand, the minimum average payoff 
per sites coincides with the maximum of $D$'s concentration. 

\begin{figure}
\centerline{\epsfxsize=7cm
            \epsfbox{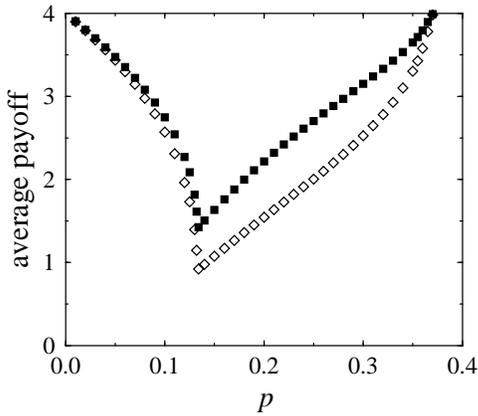}
            \vspace*{1mm} }
\caption{Average payoff per site sites versus $p$ in the two-dimensional
model. The closed squares represent the MC data obtained for $b=1.83$
and $L=512$. The diamonds indicate the quantity $4(c_C+c_T)$.}
\label{fig:PAYOFF}
\end{figure}

In Fig.~\ref{fig:PAYOFF} the  average payoff per site is  compared with 
the quantity $4(c_C+c_T)$. One can see that the agreement between these two 
quantities is generally reasonable and is even quite good for $p <p_{c1}$.
The differences between the two quantities are only coming from the fights 
taking place at the boundaries $D$-$C$ or $D$-$T$. 
However, for this range of values of $p$, the $D$ players form clusters 
due to the presence of the $T$'s and accordingly these fights are not
frequent.

Finally, let us briefly consider the case for which the parameter $b$ belongs
to the second possible range, $1<b<3/2$. The results of the simulations are
qualitatively similar to the case $3/2<b<2$. The critical values of $p$
are lower, namely, $p_{c1}=0.112(1)$ and $p_{c2}=0.297(1)$. The most
relevant differences can be observed in the limit $p \to 0$, where the
maximum concentration of the $T$ strategy is strikingly lower ($c_T <
0.15$) than those reported in the previous case (see Fig.~\ref{fig:CP2D}).

\subsection{Three-dimensional system}
\label{subsec:3D}

We now consider the case in which the players are located on the sites of
a three-dimensional cubic lattice. According to the payoff matrix (see
Table 1), five ranges of values of $b: 1<b<2$ have now to be distinguished,
separated by the following values: $b=5/4$, $4/3$, $3/2$, and $5/3$.
Any value of $b$ in one of those ranges will lead strictly to the same 
behavior.
Moreover it turns out that all the values of $b: 1 \le b\le 2$ leads 
qualitatively to the same behavior.

\begin{figure}
\centerline{\epsfxsize=7cm
            \epsfbox{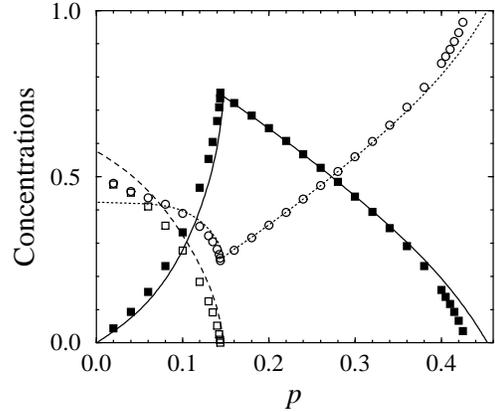}
            \vspace*{1mm}       }
\caption{Stationary state densities of $D$ (closed squares), $C$ (open
circles) and $T$ (open squares) strategies as a function of $p$ as
obtained by MC simulation of a system of size $80 \times 80 
\times 80$. The solid ($D$), dotted ($C$) and long-dashed ($T$) lines
indicate the predictions of the pair approximation.}
\label{fig:CP3D}
\end{figure}

Figure \ref{fig:CP3D} shows a typical phase diagram which is qualitatively 
similar to the one found for the two-dimensional model.
MC simulations have been performed for systems of sizes
$32^3$, $64^3$ and $80^3$
for five representative values of $b$ exploring all the different ranges.

When the value of $p$ increases, the  system undergoes
two subsequent phase transitions. The critical values of $p$ vary weakly
with  $b$. For example, for $3/2 < b <2$, we have obtained
$p_{c1}^{(MC)}=0.1441(1)$ and $p_{c2}^{(MC)}=0.4292(5)$, while for
$1<b<3/2$, we found $p_{c1}^{(MC)}=0.1512(2)$ and $p_{c2}^{(MC)}=0.4130(3)$. 
As expected, the vanishing concentrations behave near
the critical points as a power law and both transitions belong to the
directed percolation universality class. For $b=1.6$, the numerical analysis
of the MC data in the vicinity of the second transition gives
$p_{c2}^{(MC)}=0.4165(2)$ and $\beta_2=0.79(4)$ in good agreement with
the corresponding directed percolation value of $\beta=0.81(2)$ \cite{exp3d}. 

The analysis of the concentration fluctuations $\chi_T$ and $\chi_D$ 
in the vicinity of the points $p=p_{c1}$ and 
$p=p_{c2}$ leads to the critical exponent  $\gamma=0.18(8)$. 
The large uncertainty is due to the small extension of the critical 
regime (typically $\vert p-p_c \vert \le 10^{-3}$). 
The above value of $\gamma$ is compatible with the scaling law for the DP
exponents.

As far as mean-field like approximations are concerned, they are supposed to 
be better when increasing the number of nearest neighbors (dimensions).
The algebra becomes soon very cumbersome therefore our analysis is
restricted to the pair approximation. 
The results are given in Fig.~\ref{fig:CP3D}. One can see that the
approximate results are in good agreement with the MC data but in the
close vicinity of the critical points.
Note finally that the $p$-dependence of the average payoff is 
qualitatively similar to those found for the two-dimensional model 
(see Fig.~\ref{fig:PAYOFF}).

\section{CONCLUSIONS}
\label{sec:conc}

We have studied quantitatively the emergence of cooperation in a
spatially extended version of the prisoner's dilemma game with three
possible strategies (cooperation, defection and Tit for Tat) in the
absence and presence of externally enforced cooperation. The players
are distributed on a $d$-dimensional simple cubic lattice and their
interactions are restricted to nearest neighbors.

In the absence of external constraint the time evolution is controlled by a
local adoption of a neighboring strategy whose introduction is motivated
by the Darwinian selection rule. 
When starting the simulations with a random initial state 
made of only cooperators and defectors, one founds that, both in  $d=1, 2$ 
and $3$ dimensions, the stationary state of the system is a pure defector one.
However, if the $T$ strategy is also present in the initial state, then the
stationary state is dominated by mutual cooperation.

The external constraint, forcing the players to adopt the strategy $C$ with 
probability $p$, has the following consequences. If only the $C$ and $D$
strategies are present in the initial state, then the 
external constraint enforces the cooperation for all value of $p$. However, 
if the three strategies $C, D$ and $T$ are initially present and if the
dimensionality of the system is larger than one, then the external 
constraint reduces the cooperation maintained by $T$ for small values of
$p$.
The cooperation will be enforced only if the constraint is large enough
($p > p_{c1}$). These conclusions are reached both from the extended
mean-field analysis and the MC simulations for  $2$ and
$3$ dimensional systems. The general features are not affected by the value of
$b$ characterizing the temptation to defect.

Our study confirms the crucial role of the $T$ strategy which is able
to prevent the spreading of defection. The $T$ strategy, however, dyes
out in the one-dimensional system as well as in 
the models for which the simple mean-field theory is exact.

The above conclusions are in agreement with several historical facts coming 
both from the political or economical world. For example, it shows that 
forcing a fraction of the population to cooperate in a naive way 
($C$ strategy) does not improve the overall cooperation in the society.
It is better to educate more individuals in such a way that they will
be able to play the more sophisticated Tit for Tat strategy if one
desires to improve the cooperation.

From a nonequilibrium phase transition point of view, the above
investigations have confirmed that the two second order phase transitions 
associated with the  extinction processes belong to the robust directed 
percolation universality class. 

Finally we emphasize that similar behavior is expected for spatially extended
Lotka-Volterra like systems with three (or more) species providing that 
where one of the species is externally favored.

\acknowledgements

We thank Gunter Sch\"utz for a very helpful discussion.
This work was supported by the Hungarian National Research Fund
under Grant No. T-23552 and by the Swiss National Foundation.

\end{document}